\documentclass[journal,10pt]{IEEEtran}
\usepackage{textcomp}
\usepackage{times,graphicx, amsfonts}
\usepackage[cmex10]{amsmath}
\usepackage{algorithmic}
\usepackage{array}
\usepackage{algorithm,color}
\usepackage{url}
\usepackage{float}
\usepackage[T1]{fontenc}
\usepackage{amssymb}
\usepackage{cite}
\usepackage{mathrsfs}
\usepackage[centerlast]{caption}
\usepackage{flushend}

\begin{document}
\title{Beyond Powers of Two: Hexagonal Modulation and Non-Binary Coding for Wireless Communication Systems}

\author{Zhe Yang$^1$, Lin Cai$^2$\thanks{
Corresponding Author: Prof. Lin Cai, Dept. of Electrical and Computer Engineering, University of Victoria, Victoria, BC V8W 2Y2, Canada, Email: cai@ece.uvic.ca.}, Aaron Gulliver$^2$, Liang He$^3$ and Jianping Pan$^4$\\
$^1$School of Computer Science and Engineering, Northwestern Polytechnical University, Xi'an, China\\
$^2$Dept. of Electrical and Computer Engineering, University of Victoria, Canada\\
$^3$ University of Michigan, Ann Arbor, MI, USA\\
$^4$Dept. of Computer Science, University of Victoria, Canada}
{}
\maketitle

\begin{abstract}
Adaptive modulation and coding~(AMC) is widely employed in modern wireless communication systems to improve the transmission efficiency
by adjusting the transmission rate according to the channel conditions. Thus, AMC can provide very efficient use of channel resources especially over fading channels.
Quadrature Amplitude Modulation (QAM) is an efficient and widely employed digital modulation technique.
It typically employs a rectangular signal constellation.
Therefore the decision regions of the constellation are square partitions of the two-dimensional signal space.
However, it is well known that hexagons rather than squares provide the most compact regular tiling in two dimensions.
A compact tiling means a dense packing of the constellation points and thus more energy efficient data transmission.
Hexagonal modulation can be difficult to implement because it does not fit well with the usual power-of-two symbol sizes employed with binary data.
To overcome this problem, non-binary coding is combined with hexagonal modulation in this paper to provide a system which is compatible with binary data.
The feasibility and efficiency are evaluated using a software-defined radio~(SDR) based prototype.
Extensive simulation results are presented which show that this approach can provide improved energy efficiency and spectrum utilization in wireless communication systems.
\end{abstract}


\section{Introduction}
Wireless communications have become an essential component of modern information systems.
According to the Cisco Visual Networking Index and numerous studies based on market status and trends,
mobile data traffic has increased 18-fold from 2011 to 2016~\cite{index2012global}.
This ever-growing demand for mobile data creates significant demands on the scarce wireless spectrum and limited power of mobile devices.
Thus, improving the spectral and energy efficiency of wireless communication systems is an important challenge for the research community and industry.

Wireless channels typically suffer from path-loss and the effects of multipath fading and shadowing, resulting in wide variations in received signal quality.
As a consequence, many modern digital communication systems employ adaptive modulation and coding (AMC) to adjust the transmission rate according
to the time-varying channel conditions.
AMC can efficiently utilize the available bandwidth while meeting the bit-error-rate~(BER) requirements.
Quadrature amplitude modulation (QAM) is widely employed to transmit more than one bit per modulation symbol.
The most common QAM signal constellations are QPSK and 16-, 64- and 256-QAM, which carry 2, 4, 6 and 8 bits per symbol, respectively~\cite{jablon1992joint}.

QAM demodulation converts the received signal, which may be affected by fading, noise and interference, to bits.
The signal space is partitioned into decision regions for this purpose, and demodulation is achieved by determining the region that contains the received signal.
With conventional QAM, the signal space is partitioned into rectangular decision regions;
however, it is well known that a two-dimensional regular tiling with hexagons is the most efficient packing in terms of compactness~\cite{1986tilings}.
Therefore, if hexagonal decision regions are employed to partition the signal space, referred to as hexagonal quadrature amplitude modulation~(H-QAM),
the spectrum and/or energy efficiency can be improved.
H-QAM maximizes the minimum distance between signals in the constellation
and thus minimizes the symbol error probability for a given average signal energy as well as
the peak-to-average power ratio,
which is important for OFDM systems~\cite{Tanahashi09hex, Han06hex,murphy2000high,engdahl1998comparison}.
In~\cite{xu2016max}, a hexagonal lattice was employed in the time-frequency domain to enhance system performance.
However, there has been little interest in H-QAM because of the inherent difficulty in using H-QAM with binary data.
The number of constellation points may not be a power-of-two, while existing information systems are based on binary data.
A number of approaches have been proposed to overcome this problem~\cite{Tanahashi09hex, murphy2000high}.
One solution is to convert the binary data stream to non-binary symbols, e.g. using a binary-input and ternary-output (BITO)
code at the transmitter and reverse the procedure at the receiver~\cite{Tanahashi09hex, Han06hex}.
Another approach is to leave some symbols unused~\cite{murphy2000high}.
Both of these techniques cannot fully utilize the gains possible with H-QAM, especially when the size of the signal constellation is small.
Thus, current H-QAM solutions alone cannot provide sufficient performance improvement for wireless communication systems,
although it has been adopted in optical systems~\cite{wang2015generation}.

To efficiently explore the potential of H-QAM for wireless communication systems,
we propose to go beyond the conventional binary bit-mapping and coding.
Thus, in this paper we employ H-QAM with ternary digits (trits).
Ternary architectures have previously been considered in computing and storage systems due to their higher radix economy and the three usable states for
certain electromagnetic materials~\cite{krueger1995ternary,krishnan2009error}.
Although ternary communication and computing systems have not yet reached commercial
viability, their future use has been predicted by Knuth~\cite{Knuth}.

The main contributions of this paper are as follows.
\begin{enumerate}
\item  New H-QAM modulation schemes are proposed based on hexagonal tiling and the corresponding BER performance is evaluated.
These new schemes contain 3, 6, 8, and 12 constellation points to represent 1 trit, 1 bit plus 1 trit, 3 bits,
and 2 bits plus 1 trit, respectively.

\item Ternary convolutional coding is used to protect the trits directly.
For H-QAM with hybrid bit and trit information, we consider a combination of binary and ternary coding.
The BER performance is evaluated for different modulation schemes, including conventional rectangular QAM,
with code rates 1/2 and 3/4 to conform to the IEEE 802.11 standard~\cite{o2005ieee}.
These results show that the the new modulation and coding schemes not only provide finer granularity adjustment for AMC, but also can replace
some of the existing schemes by achieving a higher throughput with a lower BER for the given SNR region.

\item A prototype H-QAM wireless communication system is presented which employs non-binary information mapping using GNU Radio and USRP2,
a commonly used software-defined radio~(SDR) platform~\cite{ettus2008ettus}.
To the best of our knowledge, this is the first H-QAM based prototype communication system which demonstrates the feasibility and efficacy of hexagonal signal constellations with non-binary coding.
Further, extensive simulation results are presented to demonstrate the efficiency of the proposed scheme, which can provide considerable
performance gains compared to existing binary systems.

\end{enumerate}


In summary, this paper demonstrates the efficiency of H-QAM with non-binary symbol
mapping and error control coding in wireless communication systems.
This deviates from conventional approaches that employ rectangular constellations with binary coding.

The remainder of this paper is organized as follows.
In Section~\ref{sec:backgound}, we discuss the background and related work.
Section~\ref{sec:design} presents the non-binary communication system, including the hexagonal symbol
constellation structure, information-to-symbol mapping, non-binary
error correction coding and interleaving, and the packetization interface with the upper layers.
The system performance is investigated in Section~\ref{sec:performance} through extensive trace-driven simulation,
and the prototype system and measurement results are described in Section~\ref{sec:prototype}.
Finally, some conclusions are given in Section~\ref{sec:conclusion} along with suggestions for future work.

\section{Background and Related Work}
\label{sec:backgound}


\subsection{Hexagonal Signal Constellations}\label{subsec:hex_cons}

Modulation is the process of converting a data stream to waveforms suitable for transmission through a communication channel by varying one or more of the waveform properties,
e.g. amplitude, phase, or frequency.
For bandwidth limited bandpass modulation, quadrature amplitude modulation (QAM) is commonly employed.
Typical QAM constellations can be considered as rectangular partitions of the two-dimension signal space.
It is well known that regular hexagons provide the densest two-dimension packing, and this has motivated
research into the potential of H-QAM~\cite{yang2014hierarchical,Han06hex,Tanahashi09hex,murphy2000high,han2008use,hosur2013hexagonal,Nakamura2002Ternary}.
Existing work on hexagonal modulation usually considers one of the following approaches.
The first considers binary data so that each modulation symbol represents an integer number of bits.
However, this scheme requires that the number of constellation points be a power-of-two.
Since the number of H-QAM constellation points is  not  a power-of-two \cite{Han06hex},
some of the constellation points are not used, which is a waste of available resources.

The second approach uses all points in the hexagonal constellation for transmission to maximize the per-symbol throughput.
It has been shown in~\cite{Tanahashi09hex} that hexagonal-18 QAM
modulation (H18-QAM) requires less energy per-bit than 16-QAM.
As $18=2\times 3\times 3$, one H18-QAM symbol can be decomposed into one bit and two trits.
To accommodate a binary data stream, binary symbols can be mapped to
ternary symbols using binary-input ternary-output (BITO) convolutional or turbo codes to also provide error correction~\cite{Tanahashi09hex}.
However, this conversion is not flexible and suffers from poor performance since it fails to offer adequate protection for the ternary digits.
According to the simulation results, the coded error performance of H18-QAM is $0.6$~dB lower than that of the coded 16-QAM,
so in this case hexagonal modulation provides no improvement over rectangular modulation.

From the existing literature, it can concluded that
using binary data with non-binary modulation may be suboptimal as some constellation points are not used, and
employing BITO codes for error correction leads to inflexibility in multiplexing bits and trits.
An alternative approach is to combine H-QAM with non-binary, in particular ternary, coding, which is advocated in this paper.
The effectiveness of H-QAM in multi-media transmission using H-QAM to transmit layered video has previously been demonstrated~\cite{yang2014hierarchical}.

\subsection{Ternary Computing and Communications}
Radix economy is used to measure the cost of storing or transmitting numbers in a given base~\cite{Hayes01third}.
 It is defined as the number of digits needed to represent a number $N$ in base $b$ multiplied by the radix $b$.
 A base with a lower radix economy has a higher efficiency.
It has been proven that a radix of three provides the lowest radix economy among all integer bases.
This implies that a ternary base outperforms the widely used binary base in terms of both storage and communications.
Ternary based data recording/storing systems have been investigated~\cite{krishnan2009error,krueger1995ternary},
and the the cost of storing numbers can be minimized if a ternary base is used~\cite{Hayes01third}.
Computers using balanced ternary logic were implemented in the late 1950s and were shown to be more efficient compared with the binary based computers.
Knuth has predicted that the elegance and efficiency of ternary logic will result in its emergence in the future~\cite{Knuth}.


As discussed in Section~\ref{subsec:hex_cons}, the use of ternary codes with hexagonal constellations can fully realize the potential of H-QAM
and improve the performance of wireless communication systems.
Thus, we have conducted an extensive literature survey to find the best known ternary convolutional codes,
and these codes are punctured to obtain different code rates.
The details will be given in Section~\ref{sec:ternarycode}.



\subsection{Modulation vs.  Coding Gain}

\begin{figure}
    \centering
    \includegraphics[width=0.5\columnwidth, angle=270]{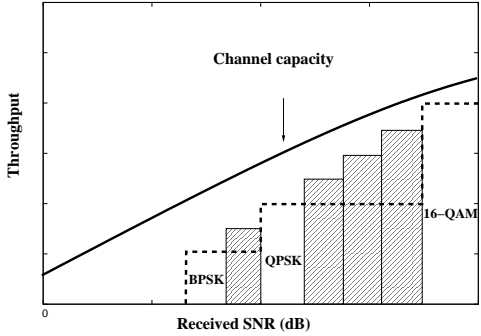}
    \caption {The achievable throughput with different modulation schemes.}
    \label{fig:shannon}
\end{figure}

The motivation of AMC is to combine different modulation schemes and code rates to fully utilize the capacity of the time-varying channel.
The code rate can be adjusted to correct different numbers of bits in error according to the BER after demodulation~(uncoded BER).
If this BER is below a given threshold, coding can be employed to reduce it to a negligible level.
Otherwise, it can be very difficult to reduce the BER to an acceptable level even with a powerful coding scheme,
and the BER with and without coding may actually be similar.
To overcome this issue, for a given SNR at the receiver, an appropriate modulation scheme
can be used to ensure that the uncoded BER after demodulation is below a desired threshold (often $10^{-3}$ for wireless
systems), and then error control coding can be applied to further reduce the BER to an acceptable region (e.g., below $10^{-6}$).
As error control coding may not be effective unless the uncoded BER is sufficiently low,
a better modulation scheme can improve the overall system performance.
For wireless communication systems employing AMC, several combinations of modulation and coding schemes are usually adopted to fit different channel conditions.
In this paper, we also consider different code rates for H-QAM.
Introducing H-QAM modulation
with coding can outperform and thus replace some of the existing rectangular QAM based AMC schemes.

In addition, if the SNR gap between two AMC schemes is large, e.g., QPSK with a rate 3/4 code and 16-QAM with a rate 1/2 code as in 802.11 standard~\cite{gast2005802},
this gap can be filled with an H-QAM based scheme.
As illustrated in Figure~\ref{fig:shannon}, since the SNR at the receiver is continuous,
there is space (shaded areas) for new modulation schemes (using new or existing coding schemes) to improve the system performance,
which motivates the work reported in this paper.

\section{System Design}
\label{sec:design}

As mentioned in the previous section, the number of constellation
points in the most compact hexagonal constellations is not always an integer power-of-two.
Therefore, to fully utilize these constellations for modulation, both bits and trits should be transmitted.
This requires a reinvestigation of the modulation constellation geometry, the mapping of bits and trits to constellation points,
the error control coding, and the multiplexing of bits and trits.

\subsection{Constellation Geometry}

A signal can be represented in the signal space domain using an in-phase and quadrature-phase (I/Q) constellation diagram.
For a constellation with $N$ points, the information carried in each symbol equals $\log_2 {N}$ bits.
The distance between a constellation point to the origin, $d$, is proportional to the square root of the transmitted symbol energy.
In the absence of noise, the received signal constellation has the same
shape as the transmitted constellation, except that, at the
receiver, the distance from a constellation point to the
origin is proportional to the square root of the received symbol energy.
In the following, {\em constellation} refers to the constellation at the receiver unless otherwise stated.

In an additive white Gaussian noise (AWGN) channel, a received symbol
follows a two-dimensional Gaussian distribution centered at the corresponding constellation point.
A Voronoi diagram can be used to determine the decision boundary of each symbol.
The probability that a symbol is demodulated in error is equal
to the probability that the received symbol lies outside the decision region of the intended symbol.

Given the fact that the Gaussian distribution decays exponentially and the BER after demodulation
should be sufficiently low (e.g., below $10^{-3}$), the
symbol error probability for H-QAM can be approximated as
\begin{equation}
\label{qfunction}
SER=2 Q\left(\sqrt{\frac{2 r^2}{N_0}}\right),
\end{equation}
where $Q(\cdot)$
is the Q-function, $r$ is the shortest distance from the constellation
point to its decision boundary, and $N_0$ is the noise spectral density.
As $r$ is equal to half of the minimum Euclidean distance in the signal space between
two constellation points, the BER is determined by the minimum Euclidean distance between constellation points.

A good modulation constellation should convey more information
under the same average power (symbol energy) and BER constraints.
Using (\ref{qfunction}), this can be converted to the following geometry problem.
In a circle of radius $\sqrt{E}+r$, pack as many non-overlapping circles with radius $r$ as
possible, where $E$ is the maximum received symbol energy and $r$ is determined by the BER constraint.

For a sufficiently large radius, $\sqrt{E}$, the optimal packing is a hexagonal tiling.
Comparing this tiling with the rectangular tiling widely used in existing QAM schemes, a hexagonal
tiling can cover the region more efficiently, which leads to approximately a $0.6$~dB gain~\cite{forney84}.

\subsubsection{The Proposed H-QAM Geometry}

In this paper, we propose TPSK, H6-QAM, H8-QAM, and H12-QAM schemes as design examples to
demonstrate the benefits of hexagonal modulation combined with ternary coding.
%
Conditioned on maximizing the minimum Euclidean distance, the {\em average}
symbol energy should be minimized.
This is equivalent to minimizing the average of the squares of the distances from the constellation points to the
origin under the condition that the minimum distance of each
constellation point to its decision boundary is no smaller than $r$, i.e.
\begin{equation}
\min \frac{\sum_i d_i^2}{M},
\end{equation}
where $d_i$ is the distance of the $i$-th constellation
point to the origin and $M$ is the number of constellation points.
Thus, the points should be as close to the origin as possible. 

\begin{figure}[!]
    \centering
    \includegraphics[width=0.9\columnwidth, angle=0]{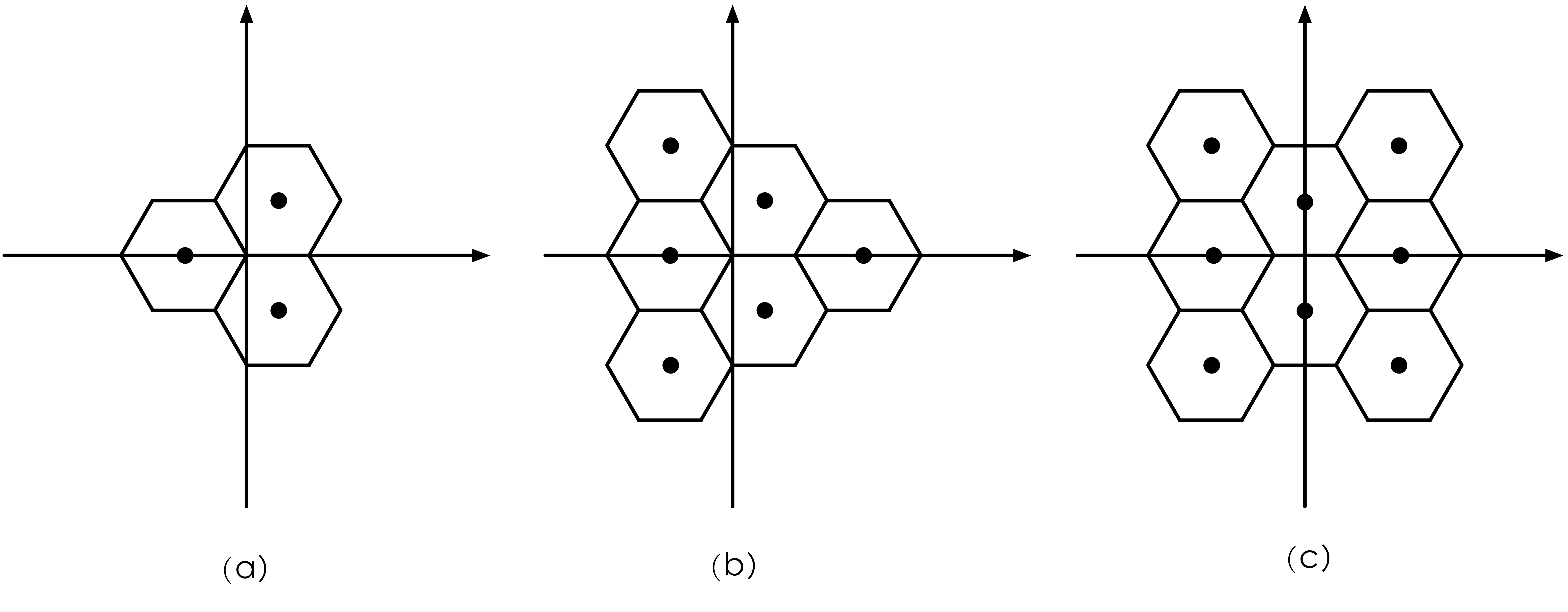}
    \caption {TPSK, H6-QAM, and H8-QAM constellations.}
    \label{fig:mapping-368}
\end{figure}

\begin{figure}
    \centering
    \includegraphics[width=0.9\columnwidth, angle=0]{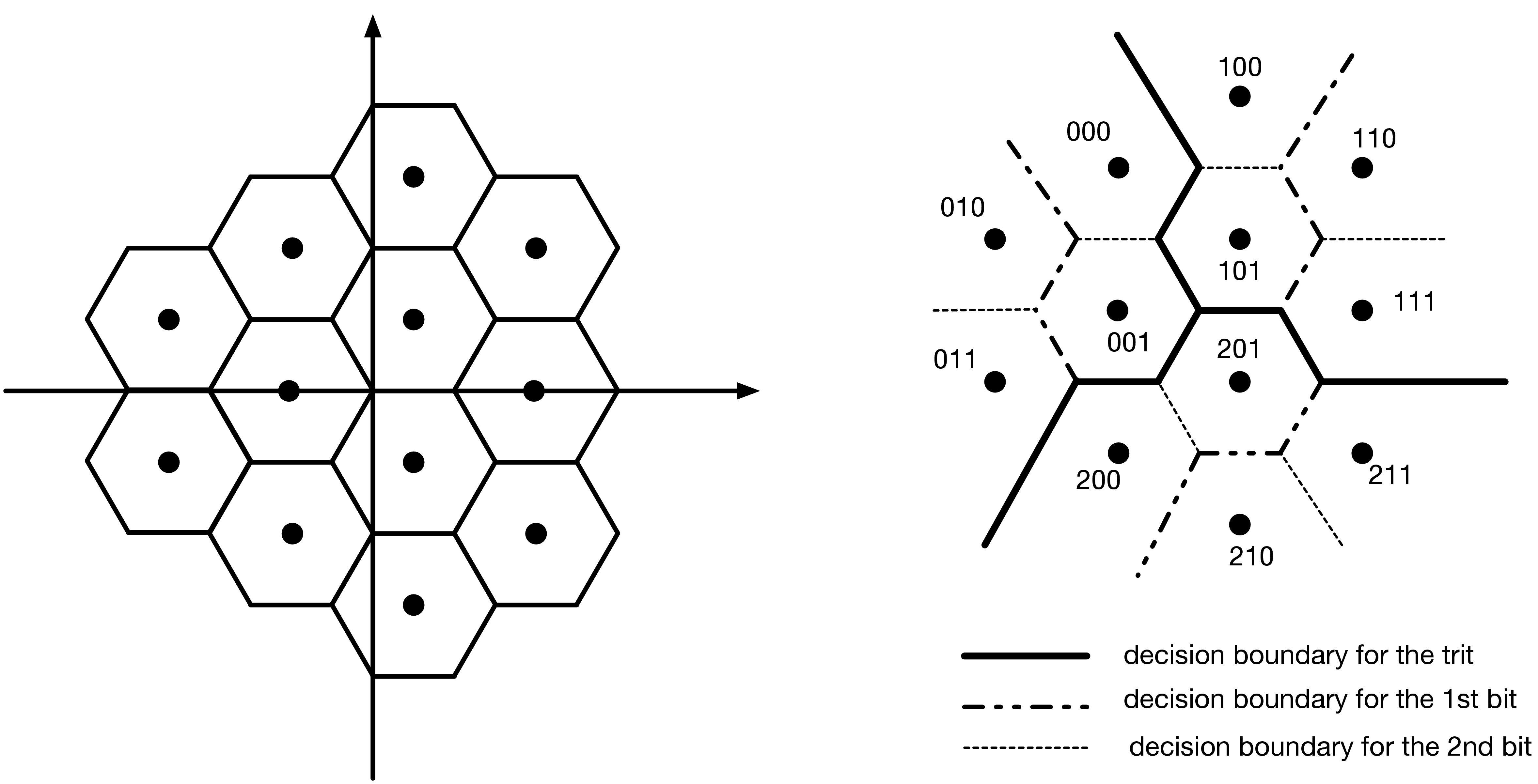}
    \caption {Constellation and mapping for H12-QAM.}
   \label{fig:mapping-12}
\end{figure}

For the TPSK constellation shown in Figure~\ref{fig:mapping-368}~(a), each
symbol represents one trit, or $\log_2 3 \approx 1.585$ bits of information.
It is straightforward to arrange the constellation
points as the end points of  an equilateral triangle, and set the origin to the center of the triangle.
The minimum Euclidean distance from each point to its decision boundary is then $r=\sqrt{3E_s/4}$,
where $E_s$ equals the average symbol energy.
%
For H6-QAM, each symbol carries one trit plus one bit of information, and
the constellation arrangement is shown in Figure~\ref{fig:mapping-368}~(b).
The minimum Euclidean distance from a constellation point to its decision boundary is $r=\sqrt{8E_s/15}$,
which is much larger than that of 6-PSK ($\sqrt{E_s/4}$), and thus the symbol error performance of H6-QAM is better.
Thus, H6-QAM can transmit the same amount of information as 6-PSK as they both contain $6$ constellation points, but with a lower symbol error rate.

For H8-QAM, each symbol carries three bits of information, and the
constellation arrangement is shown in Figure~\ref{fig:mapping-368}~(c),
where the origin is located at the midpoint of an edge of a hexagon.
The minimum Euclidean distance from a constellation point to
its decision boundary is $r=\sqrt{2E_s/9}$, which is larger than that of rectangular 8-QAM ($\sqrt{E_s/6}$).
For H12-QAM, each symbol carries one trit plus two bits of
information, and the proposed constellation arrangement is shown in
Figure~\ref{fig:mapping-12}~(a), where the origin is the joint vertex of the center hexagons.
The minimum Euclidean distance from a constellation point to its decision boundary is $r=\sqrt{3E_s/19}$.
Note that TPSK, H6-QAM and H12-QAM are rotationally symmetric by $120^\circ$, which provides additional benefits as will be shown in Section~\ref{sec:prototype}.

\subsection{Constellation Mapping}

Given the geometry of the constellation points, the next step is to map the bits and trits to
the constellation point such that the BER is minimized.
Compared with the conventional rectangular modulation, there are more neighboring points (with the smallest distance to a constellation
point) using hexagonal modulation, so careful mapping is required to limit the number of bit and/or trit errors due to a symbol error.
It is not straightforward to obtain a Gray type mapping (with only one bit/trit difference between neighboring points),
because the number of neighboring constellation points with hexagonal tiling often exceeds the number of digits~(bits and/or trits) represented by each symbol.

The following design principle is used here to obtain good mappings.
If starting from a bit, the constellation points are divided into two clusters.
Similarly,  the constellation points are divided into three clusters if starting from a trit.
Then, `0' and `1' (for a bit) or `0', `1' and `2' (for a trit) are arbitrarily assigned to each of the clusters.
For the remaining bits or trits, binary or ternary numbers are first assigned to
the points in one cluster, and then in turn to the points in the other clusters.
The same number is assigned to neighboring points in different clusters as much as possible.
For example, with the H12-QAM constellation in Figure~\ref{fig:mapping-12}~(b),
starting with a trit, the $12$ points are divided into three clusters, and are assigned `0', `1' and `2', respectively.
Within the first cluster, a Gray-type mapping is used to assign `00', `01', `11' and `10' to the four points.
For the second cluster, `100' and `101' are assigned to the points neighboring `000' and `001'.
Similarly, for the third cluster, `201' and `211' are assigned to the points neighboring `101' and `111'.
Although this approach to constellation mapping does not in general produce a Gray mapping,
the results presented in Section~\ref{sec:performance} show
that it still leads to a significant performance improvement.

\subsection{Ternary Error Control Coding and Interleaving}\label{sec:ternarycode}

Given a noisy channel, a received constellation point may
differ from the transmitted one, which will result in bit/trit errors.
Interleaving and error control coding can be used to mitigate these errors.

\subsubsection{Non-binary Convolutional Codes}\label{subsec:nbcc}

\begin{figure}
    \centering
    \includegraphics[width=1\columnwidth]{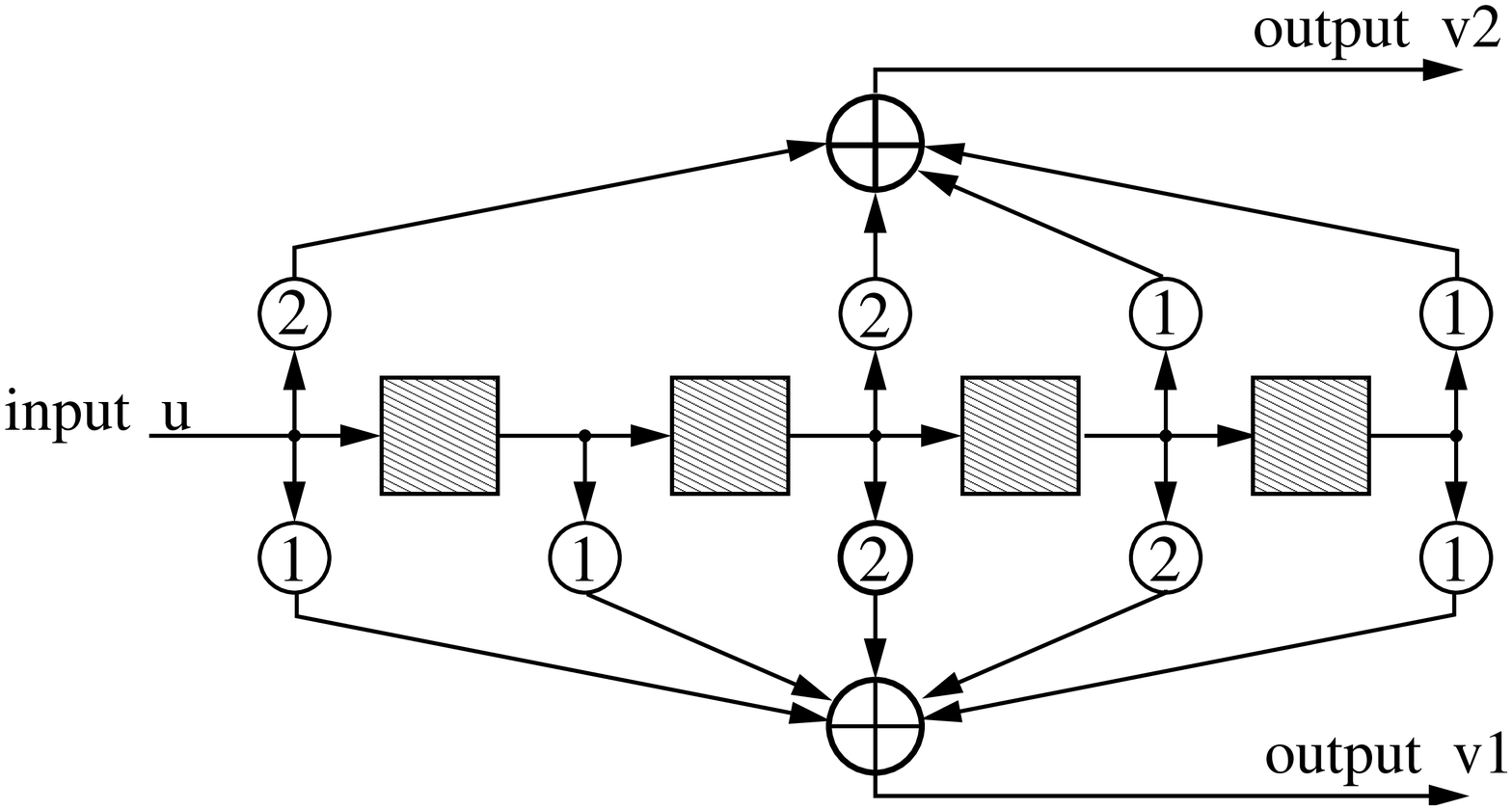}
    \caption {A ternary convolutional encoder~\cite{ternarycc}.}
    \label{fig:ternarycc}
\end{figure}

Convolutional coding has been widely used in wireless systems such as 802.11 because of the relatively simple implementation and good performance improvements~\cite{gast2005802}.
A binary convolutional encoder can be represented by the parameters $(n, k, m)$ where $k$ and $n$ are the numbers of input and output bits, respectively, and $m$ is the encoder memory size. The corresponding  code rate is $k/n$.
Three code rates, $1/2$, $3/4$ and $2/3$, are employed in the IEEE 802.11 standard.
At the receiver, the received coded bit stream is decoded to recover the original message bit stream.
The Viterbi algorithm provides maximum-likelihood decoding is widely used in practice because of the low implementation complexity and satisfactory performance.

A ternary $(n, k, m)$ convolutional encoder maps $k$ input trits to $n$ output trits~\cite{ternarycc}.
A ternary convolutional encoder can be simply implemented using shift registers and modulo-3 adders, as shown in Figure~\ref{fig:ternarycc}.
In the figure, the shaded squares are memory elements, the circled numbers are the coefficients that the trits are multiplied by, and
$\oplus$ represents a modulo-3 adder.
As there is one input trit stream $u$ and two output trit streams v1 and v2, the code rate is $1/2$.
Similar to binary convolutional codes, puncturing can be employed to obtain different code rates to adjust the protection level
according to the modulation scheme, channel condition and required BER.
Without loss of generality and to be consistent with the 802.11 standard, a rate 3/4 punctured ternary convolutional code is considered.
The puncturing pattern is the same as that used for the binary convolutional code,
$[1\hspace{0.15cm} 1\hspace{0.15cm} 1\hspace{0.15cm} 0\hspace{0.15cm} 0\hspace{0.15cm} 1]$,
where $0$ means the coded digit at that position is punctured.

The operations in a ternary Viterbi decoder are in the Galois field of three elements, $GF(3)$.
The complexity associated with a convolutional code is primarily in the decoder.
A ternary decoder using the Viterbi algorithm requires $O(3^{m}L_T)$ memory space and $O(3^{2(m+1)}L_C)$
computation time, where $L_T$ is the trace back length of the decoder and $L_C$ is the code block length, respectively.
The computational complexity of a ternary convolutional decoder is comparable to that of a binary decoder with a similar number of states.

\subsubsection{Interleaving}

Many error correcting codes such as convolutional codes cannot
tolerate burst errors, so it is desirable to separate these errors using an interleaver.
An interleaver can also help mitigate errors when Gray mapping is not employed.
%
Bits or trits are interleaved within a packet as the performance results show that
this intra-packet interleaving provides a performance gain for fading channel and does not introduce significant delay.

{\em Remark:}
Several error control coding designs exist in the literature that have been shown to provide near-capacity performance, e.g.
low-density parity check (LDPC) codes~\cite{richardson2001},
accumulate-repeat-accumulate~(ARA) codes~\cite{pfister2007accumulate},
and rateless codes~\cite{erez2012rateless}.
Further research work is required to extend these designs to ternary codes to further improve the performance of H-QAM AMC schemes.

\subsection{System Architecture}

\begin{figure*}
   \centering
   \includegraphics[width=1 \textwidth]{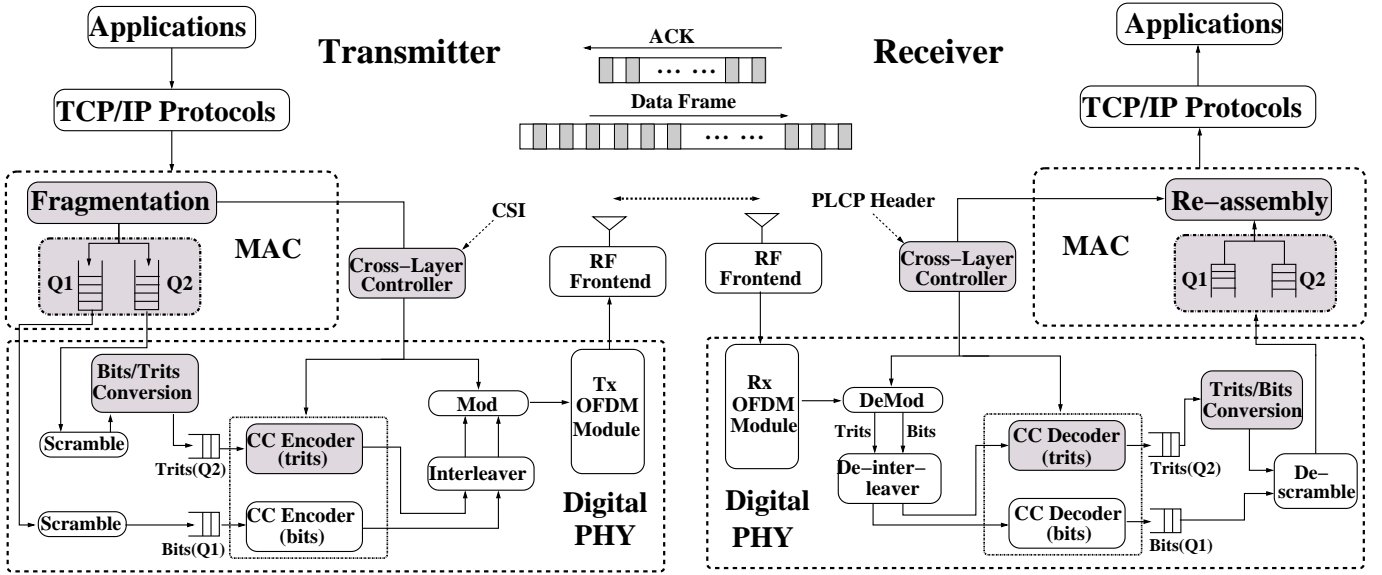}
   \caption{The new system architecture based on IEEE 802.11a. }
   \label{fig:system}
\end{figure*}

Using the IEEE 802.11a standard as an example, Figure~\ref{fig:system} shows the H-QAM system architecture.
The shaded blocks indicate new modules or existing modules that required updated.
The messages are divided into two queues, one of which is converted to
a trit stream and the other is kept as a bit stream.
The bits/trits conversion module converts the bit sequence to a trit sequence
(details of this conversion will be discussed later), and this is the input to the ternary
convolutional code~(TCC) encoder.
The cross-layer controller uses the channel state information~(CSI) to determine the modulation and
coding scheme to be used, and this is included in the Physical Layer
Convergence Procedure~(PLCP) header~\cite{gast2005802}.


The PLCP header cotains the physical layer control information.
The {\em rate} field of the PLCP header is used to inform the receiver of the modulation and coding scheme to be employed.
There are $8$ different adaptive modulation and coding~(AMC) schemes in the current IEEE 802.11a standard, which is represented by a $4$-bit {\em rate} field.
There is one reserved bit in the header which can be combined with the $4$-bit rate field to represent up to $32$ different modulation and coding schemes.
This is sufficient to include the new AMC schemes based on H-QAM and ternary convolutional coding proposed in this paper,
as some of new AMC schemes replace existing AMC schemes based on rectangular QAM and binary convolutional coding.
The proposed non-binary H-QAM communication system is compatible with conventional systems and does not require additional communication overhead as will be shown later.

\section{Performance Evaluation}
\label{sec:performance}

\subsection{Uncoded BER Performance}

\begin{figure}
    \centering
    \includegraphics[width=1\columnwidth]{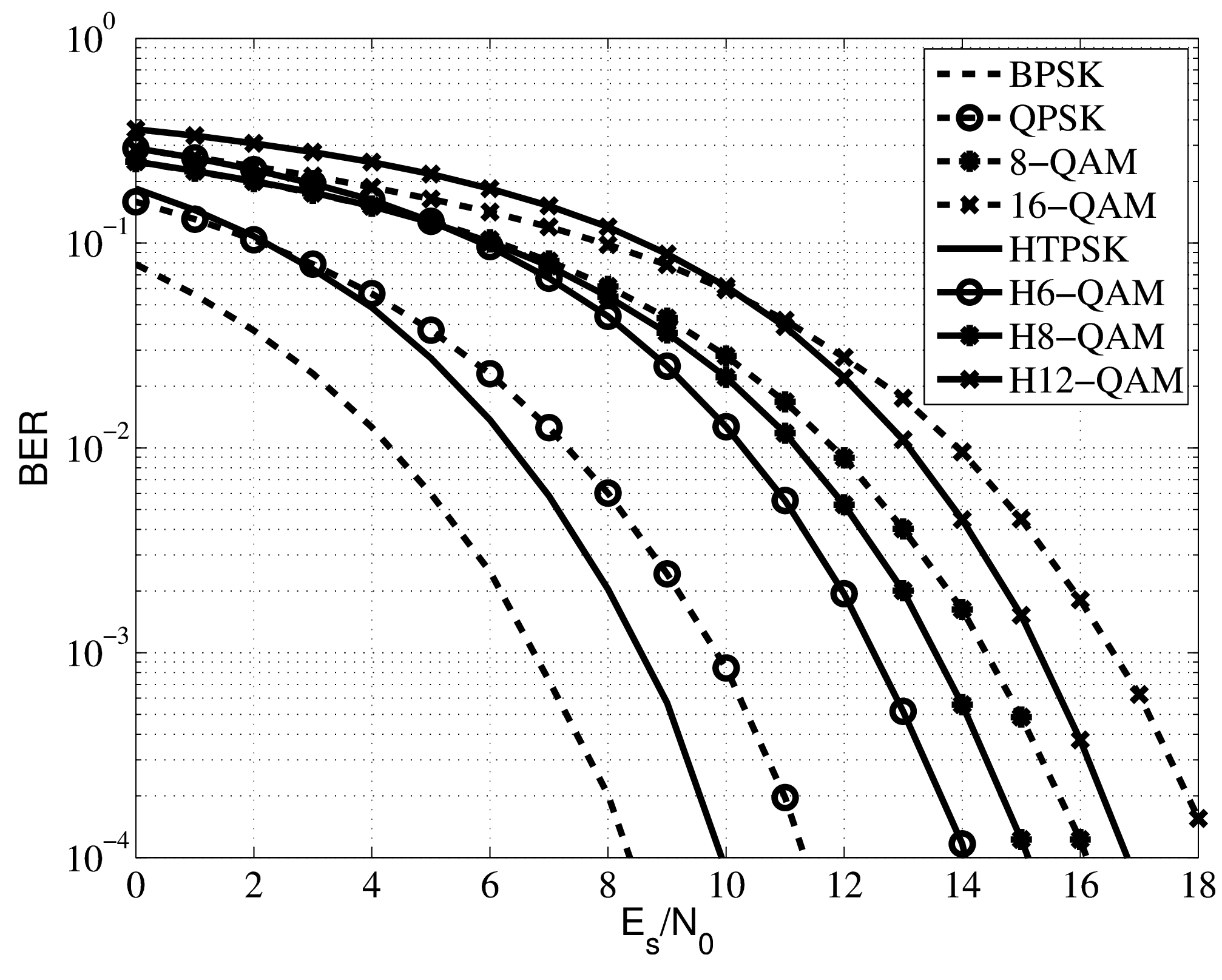}
    \caption {BER without error control coding.}
    \label{fig:rawber}
\end{figure}

A key performance indicator is the BER w.r.t the received SNR over an AWGN channel.
We consider a received SNR from $0$ to $18$~dB and examine the uncoded BER for different modulation schemes.
The received signal to noise ratio is $E_s/N_0$ where $E_s$ is the received energy per symbol and $N_0$ is the noise power spectral density.
Monte Carlo simulation with Matlab was used to obtain the BER results shown in Figure~\ref{fig:rawber}.
Each H-QAM constellation point has more neighbors than with rectangular QAM, so the resulting non-Gray code mapping will degrade the BER performance of H-QAM, but
H-QAM is still an efficient modulation design in terms of the symbol error rate.
For example, H8-QAM outperforms rectangular 8-QAM by $0.8$~dB at a BER of $10^{-3}$, and this gap increases for smaller BERs.
This demonstrates the advantage of using hexagonal constellations.

Using $10^{-3}$ as the BER benchmark, the addition of TPSK, H6-QAM, H8-QAM and H12-QAM provides more choices to adapt the modulation according to the received SNR.
For instance, TPSK can be used to replace BPSK in the SNR range $[6.2, \ 8]$~dB to achieve a $58\%$ throughput gain.
Similarly, H6-QAM, H8-QAM, and H12-QAM can replace QPSK in the SNR range of $[9.8,  \ 15.3]$~dB to achieve a $29\%$ to $79\%$ throughput gain.
Similarly, H-QAM constellations such as H27-QAM and H54-QAM should provide throughput gains when the received SNR is higher.

\subsection{Coded BER Performance}

For the coded BER performance, an AWGN channel is considered with the rate $1/2$ ternary convolutional code shown in Figure~\ref{fig:ternarycc}.
This code has $3^4=81$ states, so for a fair comparison a rate $1/2$ binary convolutional code is used with $2^6=64$ states.
This binary code is employed in the IEEE 802.11 standard~\cite{gast2005802}.
To evaluate the performance of H-QAM based AMC, we also consider punctured convolutional coding with the puncture pattern discussed in Sec.~\ref{subsec:nbcc},
which provides a code rate of $3/4$.

\begin{figure}
    \centering
    \includegraphics[width=1\columnwidth]{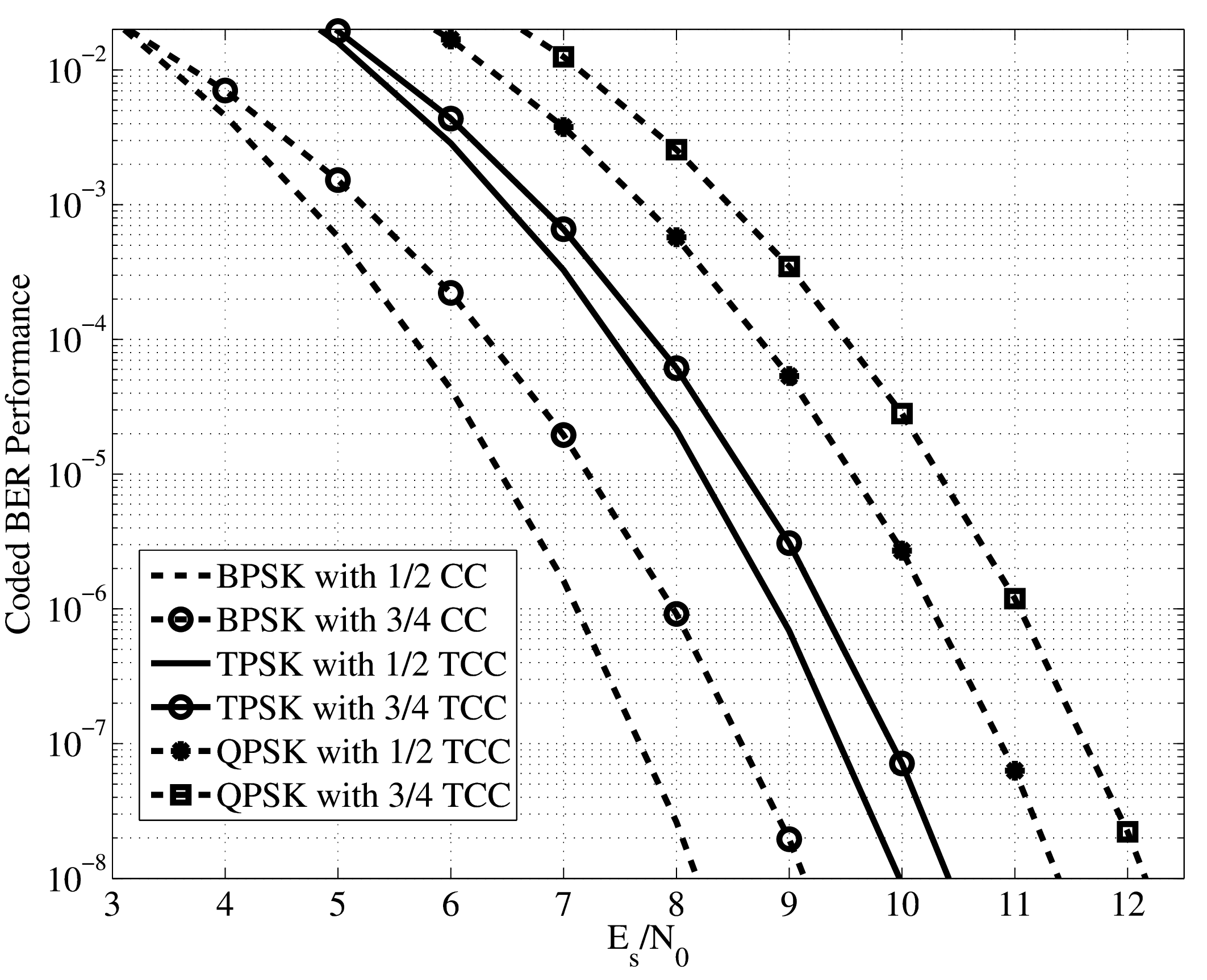}
    \caption {Coded BER with BPSK, TPSK, and QPSK modulation.}
    \label{fig:codedbtq}
\end{figure}

\begin{figure}
    \centering
    \includegraphics[width=1\columnwidth]{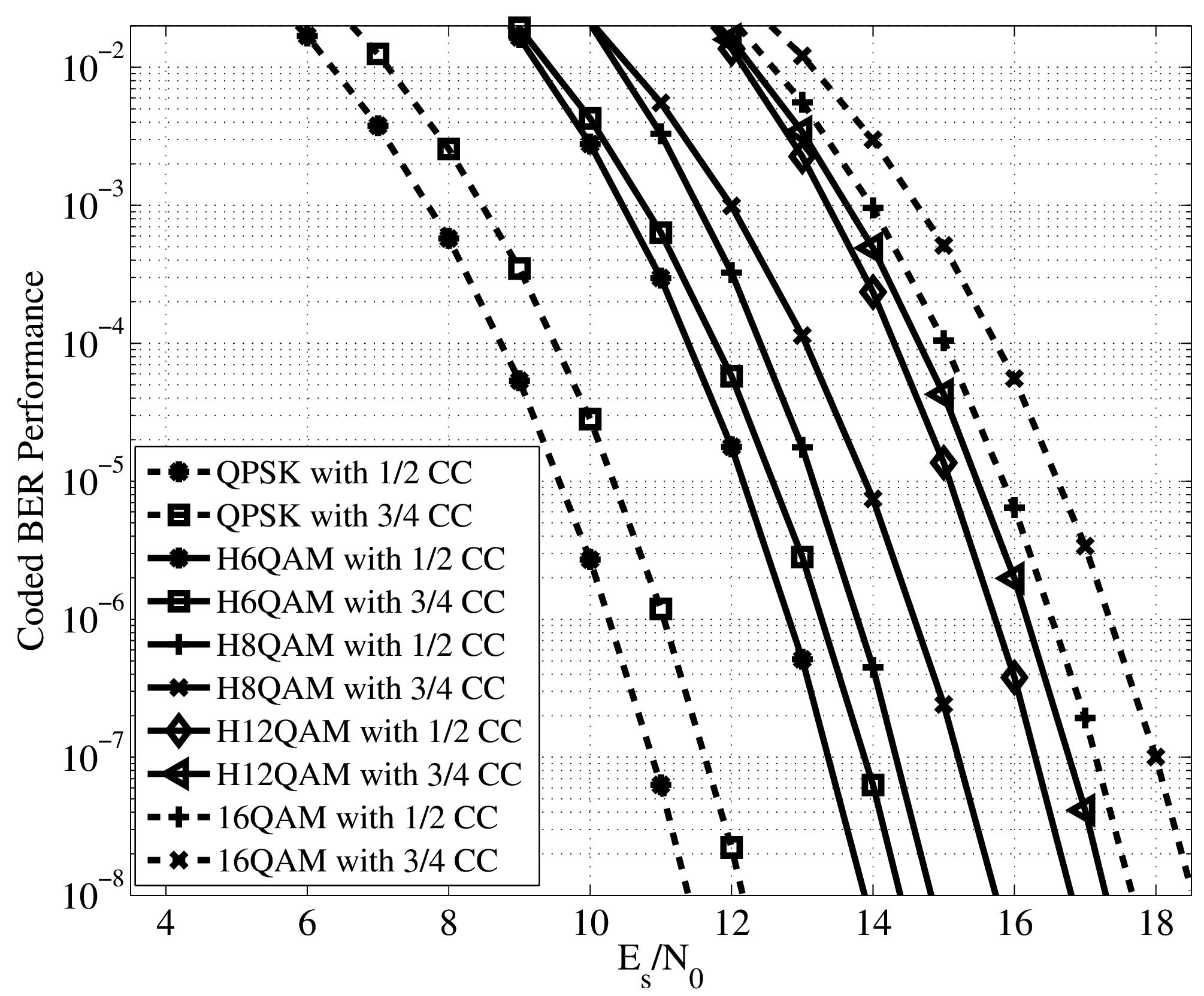}
 \caption {Coded BER with QPSK, H6-QM, H8-QAM, H12-QAM, and 16-QAM modulation.}
    \label{fig:codedother}
\end{figure}

The BERs for BPSK, TPSK and QPSK with different code rates are shown in Figure~\ref{fig:codedbtq},
and the BERs for QPSK, H6-QAM, H8-QAM, H12-QAM and 16-QAM with different code rates are given in Figure~\ref{fig:codedother}.
Comparing Figures~\ref{fig:rawber} and \ref{fig:codedbtq},
when the SNR is below $5$~dB, the BER for BPSK with or without coding is similar ($10^{-2}$ or above).
In addition, when the SNR is below $6$~dB, the BER for TPSK with or without coding is similar ($10^{-2}$ or above).
This indicates that error correction coding is effective only when the uncoded BER is sufficiently low.

Figures~\ref{fig:codedbtq} and~\ref{fig:codedother} show that for a given code rate and BER,
the required SNR increases with respect to the number of constellation points.
This is because the denser the constellation, the smaller the minimum Euclidean distance.
However, some combinations of H-QAM and coding outperform the rectangular QAM combinations in terms of both throughput (bits per symbol)
and BER. For instance, the BER performance of TPSK with a rate $3/4$ code is better than that of QPSK with a rate $1/2$ code,
with more than $0.9$~dB improvement at a BER of $10^{-6}$.  In addition,
TPSK with a rate $3/4$ code has a throughput of $1.178$~b/sym~(bits per symbol),
which is higher than that of QPSK with a rate $1/2$ code, $1$~b/sym.
Thus, H-QAM provides a $17.8\%$ throughput gain at a lower SNR, so
TPSK with rate $3/4$ coding can replace QPSK with rate $1/2$ coding in AMC.
Similarly, H12-QAM with rate $3/4$ coding can replace 16-QAM with rate $1/2$ coding.
Using $10^{-6}$ as the threshold for coded BER, the required SNR and the corresponding throughput are given in Table~\ref{tab:ts}.

\begin{table} \footnotesize
   \centering
   \caption{Comparison of Modulation and Coding Schemes}
   \begin{tabular}{|l|l|l|l|l}
    \hline
    {\bf Modulation} & {\bf Code rate}& {\bf Throughput (b/sym)} & {\bf SNR (dB)}\\ \hline
    { BPSK*} & { 1/2}& { 0.5} & { $>$ 7.12}\\ \hline
    { BPSK*} & { 3/4}& { 0.75} & { $>$ 7.97}\\ \hline
    { TPSK} & { 1/2}& { 0.785} & { $>$ 8.89}\\ \hline
    { TPSK*} & { 3/4}& { 1.178} & { $>$ 9.3}\\ \hline
    { QPSK} & { 1/2}& { 1} & { $>$ 10.2}\\ \hline
    { QPSK*} & { 3/4}& { 1.5} & { $>$ 11.04}\\ \hline
    { H6-QAM} & { 1/2}& { 1.285} & { $>$ 12.81}\\ \hline
    { H6-QAM*} & { 3/4}& { 1.928} & { $>$ 13.27}\\ \hline
    { H8-QAM} & { 1/2}& { 1.5} & { $>$ 13.78}\\ \hline
    { H8-QAM*} & { 3/4}& { 2.25} & { $>$ 14.58}\\ \hline
    { H12-QAM} & { 1/2}& { 1.785} & { $>$ 15.73}\\ \hline
    { H12-QAM*} & { 3/4}& { 2.678} & { $>$ 16.18}\\ \hline
    { 16-QAM} & { 1/2}& { 2} & { $>$ 16.53}\\ \hline
    { 16-QAM*} & { 3/4}& { 3} & { $>$ 17.35}\\ \hline

         \end{tabular}
   \label{tab:ts}
\end{table}
\normalsize

Considering AMC, TPSK and H12-QAM with rate $3/4$ coding can be used to replace QPSK and 16-QAM with rate $1/2$ coding, respectively, and
H6-QAM and H8-QAM with rate $3/4$ coding can be used in the SNR gap between QPSK with rate $3/4$ coding and 16-QAM with rate $1/2$ coding.
This provides a finer grain set of choices for AMC.
These new AMC schemes can be indicated in the PLCP header for proper demodulation and decoding at the receiver.
We next study these new modulation and coding schemes in terms of system throughput and efficiency.

\subsection{Single Link Throughput}

The new modulation and coding combinations are now considered in an AMC system.
A Rician fading channel with Rician factor $K=6$~dB is considered for a single communications link.
The conventional AMC set contains BPSK, QPSK and 16-QAM with rate $1/2$ and $3/4$ coding according to the IEEE 802.11 standard~\cite{gast2005802}.
The augmented AMC set includes the existing QAM and new H-QAM based transmission schemes marked by * in Table~\ref{tab:ts}.
For a fair comparison, we use the same symbol rate and energy for all the modulation and coding schemes.
In the simulations, data packets of size $1$ kB are transmitted, and $1,000$ packets are transmitted to obtain the average performance.
Separate intra-packet interleaving is used for the bits and trits.
The transmitter and receiver structures are as given in Figure~\ref{fig:system}.

\begin{figure}
    \centering
    \includegraphics[width=1\columnwidth]{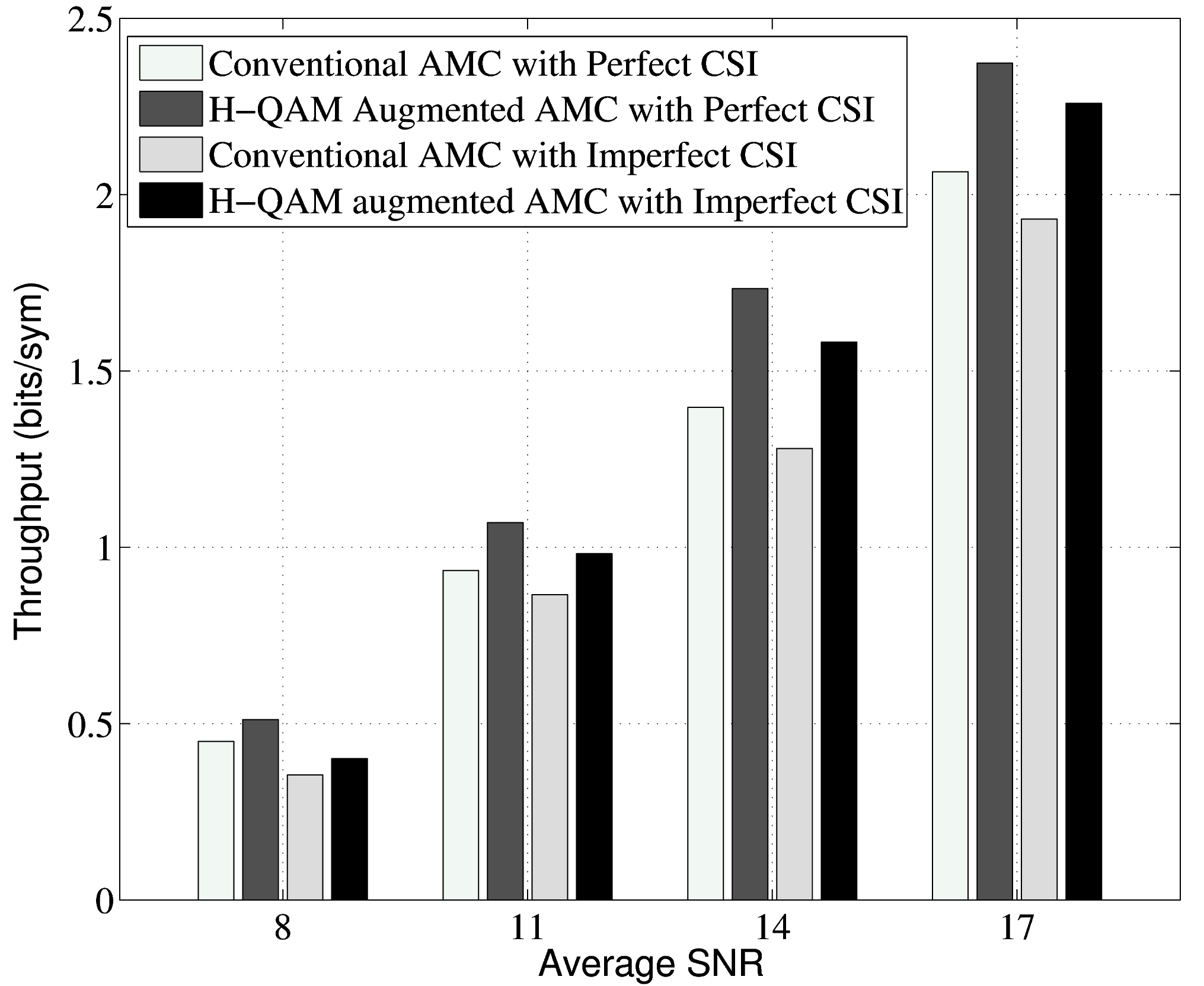}
    \caption {Single link throughput comparison.}
    \label{fig:csi-impact}
\end{figure}

A practical issue with AMC is the imperfect estimation of channel conditions.
If the received SNR is underestimated, the sender may select a modulation and coding scheme with a lower throughput (number of bits per received symbol).
Conversely, if the received SNR is overestimated, it may result in a higher BER than the required threshold, which is even more undesirable.
The impact of channel estimation errors on the system performance is thus of critical importance.
To examine this impact, channel estimation errors are modeled as a Gaussian random variable with zero mean and unit variance.
To reduce the probability that the received SNR is overestimated (which may severely degrade system performance), the transmitter
uses the estimated SNR minus its standard deviation to select the modulation and coding scheme.

Figure~\ref{fig:csi-impact} compares the link throughput using AMC with and without the proposed H-QAM schemes
for an average received SNR of $8$, $11$, $14$ and $17$~dB.
The average throughput was obtained for $1000$ Monte Carlo iterations to average the effects of fading.
With perfect channel estimation, the proposed non-binary communication system outperforms the conventional system
by $13.7\%$, $14.5\%$, $24.1\%$ and $14.9\%$ when the average SNR is $8$, $11$, $14$ and $17$~dB, respectively.
The performance of both conventional QAM and H-QAM degrades with imperfect channel estimation, but the proposed system
still achieves throughput gains of $13.1\%$, $13.5\%$, $23.6\%$ and $17\%$ for an average SNR of $8$, $11$, $14$ and $17$~dB, respectively.
These results show that the proposed H-QAM AMC can provide better BER performance and also higher throughput.
Further, it has a finer granularity than conventional QAM AMC w.r.t. the SNR.

\subsection{Network Throughput}

The system performance is now evaluated in an infrastructure-based network where an access point~(AP) is centrally located
to serve all users in the network, e.g. a WiFi network, and the mobile users are randomly distributed.
The wireless channels suffer from independent Rician block fading.
The path-loss exponent is $\alpha=3$, and all transmitted symbols have the same average energy.
We consider the downlink performance where the AP transmits packets (with size $1$ kB) to all mobile users in a round-robin manner.
It is assumed that the AP has all channel information which is used to select the AMC scheme for each packet according to the estimated SNR.
The average SNR is set to $7$~dB~when users are at the boundary of the network.

\begin{figure}
  \centering
  \includegraphics[width=1\columnwidth]{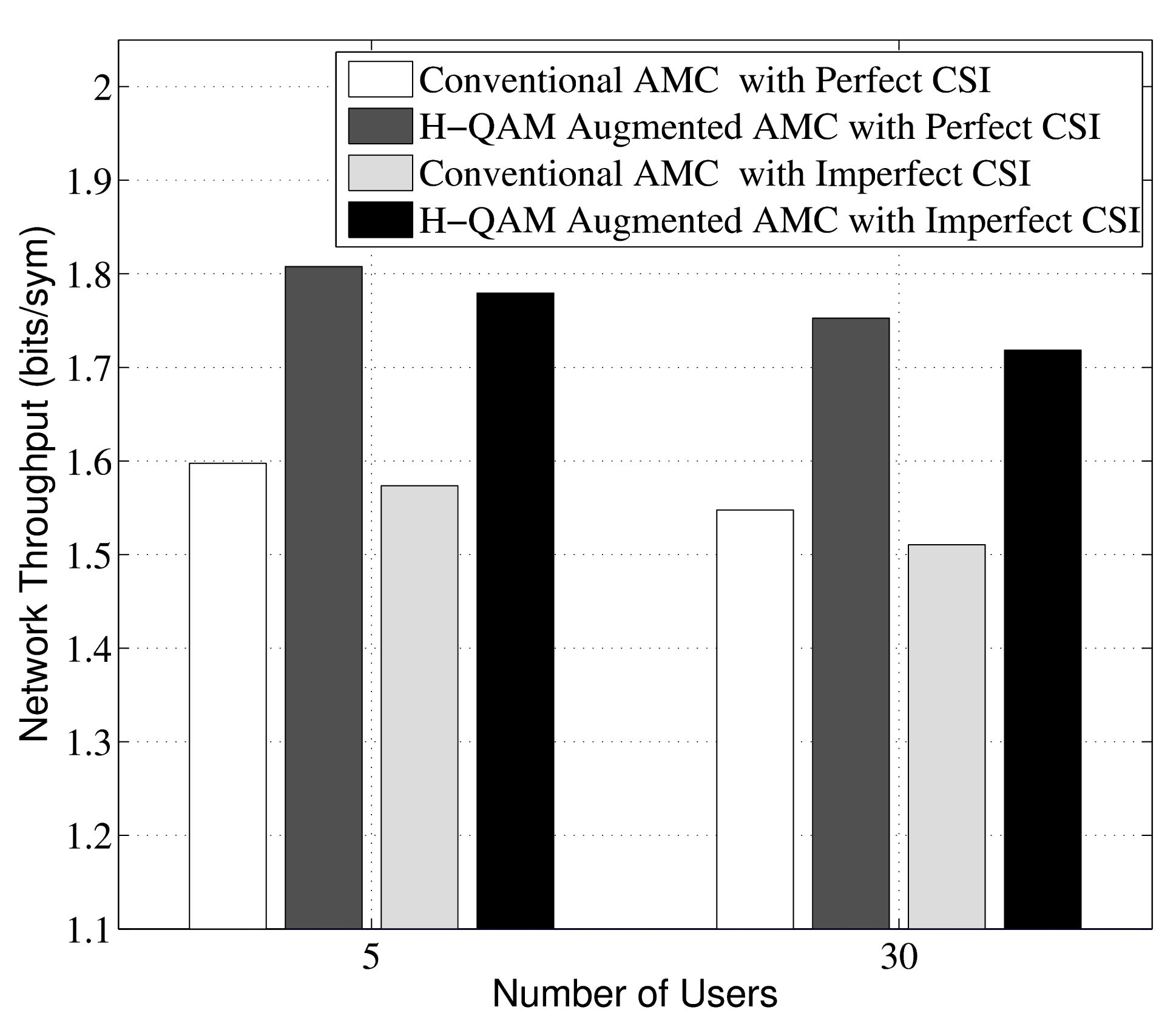}
 \caption{The network throughput with $5$ and $30$ users.}
 \label{fig:networkthroughput}
\end{figure}

The system performance was evaluated using Monte Carlo simulation with different node densities.
For each density, 1000 simulations were run using random topologies, and the average network throughput was determined in terms of bits per symbol.
Figure~\ref{fig:networkthroughput} presents the results for $5$ and $30$ users, which correspond to sparse and dense networks, respectively.
These results show that the proposed non-binary H-QAM schemes can increase the average network throughput by
more than $13.2\%$ and $13.3\%$ for the $5$ and $30$ user cases, respectively, with perfect or imperfect channel state information.
In addition to the throughput gain, for a given average transmitted symbol energy
the per bit energy is also reduced by $11.6\%$ and $11.7\%$, respectively.

\section{Prototype System and Measurements}
\label{sec:prototype}

A prototype for the non-binary H-QAM communication system was developed using the software-defined radio (SDR)
platform USRP2~\cite{ettus2008ettus} and GNU Radio.
One USRP2 was connected to the laptop host (DELL E5400) as the transmitter and another to the PC host (DELL OPTIPLEX 755) as the receiver.
%
The USRP2-based OFDM implementation~\cite{Veljko} was augmented with the proposed H-QAM.
The carrier frequency, number of subcarriers and FFT length are $2.49$~GHz, $80$ and $512$, respectively.


The three new hexagonal modulation designs, TPSK, H6-QAM and H8-QAM, were implemented
using the bits/trits conversion presented previously.
This conversion will introduce minimal overhead and thus there is a small performance loss~\cite{Tanahashi09hex}.
Using a long bit sequence will reduce this loss but increase the delay and complexity.
The conversion efficiency is defined as
\begin{equation}
\eta = {l_b\over l_t \log_2(3)},
\end{equation}
where $l_b$ is the length of the input bit sequence and $l_t$ is the length of the output trit sequence.
Converting $11$ bits to $7$ trits 
provides an efficiency of ${11\log_22 \over 7\log_23}=99.1\%$~\cite{koike2004space}.
As the block of bits is small, this can easily be implemented using a lookup table with $2^{11}$ entries.
After bits/trits conversion, the data is mapped to modulation symbols.
%

\begin{table}   \centering
\caption{The experimental BER results}
   \begin{tabular}{|l|l|l|}
    \hline
    {\bf Modulation} & {\bf Uncoded BER} & {\bf Coded BER}\\ \hline
    {BPSK} & {$5.4\times 10^{-5}$} & {$<  10^{-7}$}\\ \hline
    {TPSK} & {$5.1\times 10^{-5}$} & {$ <10^{-7}$}\\ \hline
    {QPSK} & {$2.9\times 10^{-4}$} & {$2.13\times 10^{-7}$}\\ \hline
 	 {H6-QAM} & {$1.2\times 10^{-3}$}& {$5.49\times 10^{-6}$}\\  \hline
    {H8-QAM} & {$1.6\times 10^{-3}$}& {$1.04\times 10^{-5}$}\\  \hline
    {8PSK} & {$2.3\times 10^{-3}$} & {$1.22\times 10^{-5}$}\\ \hline
         \end{tabular}
   \label{tab:measurements}
\end{table}

One thousand test frames were transmitted where each frame contains $100$ blocks of data, and each block contains $16$ bytes of data with a $4$-byte CRC.
The transmitter converts all or some of the received bits in each block to trits depending on the modulation employed.
The receiver demodulates the received symbols to bits and/or trits according to the modulation scheme used as indicated in the PLCP header.
Then the trits are converted back to bits and the CRC is checked to determine whether there are any transmission errors in the block.
If a block fails the CRC check, the number of errors is obtained by comparing it with the original block.
The BER after demodulation can then be calculated using the total number of errors.
To obtain the coded BER, the received bits or trits are further processed using binary and ternary Viterbi decoders.

The BER results are shown in Table~\ref{tab:measurements}.
As expected, the performance of TPSK is better than that of QPSK and H6-QAM, and
H8-QAM performs better than 8PSK.
These results demonstrate the feasibility and simplicity of deploying the proposed H-QAM.
Note that the uncoded BER performance of TPSK is slightly better than that of BPSK.
This is because the number of bits transmitted with a TPSK symbol is significantly higher~\cite{sklar2001digital}.
Further, the $120^\circ$ phase difference between TPSK symbols is much smaller than that of BPSK,
which is a practical benefit of TPSK modulation.
Despite the limitations of the USRP2 hardware and the effects of the fading channel,
the measured results given for an indoor environment confirm the feasibility of employing H-QAM.

\section{Conclusion}
\label{sec:conclusion}

In this paper, the design and implementation of a non-binary communication system was considered which employs Hexagonal quadrature amplitude modulation
(H-QAM) and ternary error control coding.
Both bits and trits were employed to improve the spectral and power efficiency.
Thus, the proposed system deviates from the conventional powers-of-two modulation,
although it is compatible with existing systems.
Further, the proposed H-QAM can be implemented by modifying existing QAM systems.
Four hexagonal modulation schemes were examined, but others are possible and the constellation design should depend on the particular application.
For instance, H-QAM with denser constellation points may be desirable for higher SNRs.
A ternary convolutional code with code rates $1/2$ and $3/4$ (the latter was obtained by puncturing) was employed.
The performance can be further improved if joint modulation and coding design is employed,
particularly in terms of the system capacity.
To date, the design of ternary error correcting codes has received very little attention in the literature,
so this is a promising area for further investigation.

\bibliographystyle{abbrv}

\bibliography{myb}

\end{document}